# Regional and temporal characteristics of *Bovine Tuberculosis* of cattle in Great Britain

Aristides Moustakas[1,*] and Matthew R Evans[1]

1. School of Biological and Chemical Sciences

    Queen Mary University of London

    Mile End Road, E1 4NS, London, UK

* Corresponding author

Aristides (Aris) Moustakas

Email: arismoustakas@gmail.com; a.moustakas@qmul.ac.uk

**Abstract**

Bovine tuberculosis (TB) is a chronic disease in cattle that causes a serious food security challenge to the agricultural industry in terms of dairy and meat production. Spatio-temporal disease analysis in terms of time trends and geographic disparities of disease dynamics can provide useful insights into the overall efficiency of control efforts as well as the relative efficiency of different management measures towards eradication. In GB, Scotland has had a risk based surveillance testing policy under which high risk herds are tested frequently, and in September 2009 was officially declared as TB free. Wales have had an annual or more frequent testing policy for all cattle herds since January 2010, while in England several herds are still tested every four years except some high TB prevalence areas where annual testing is applied. Time series analysis using publicly available data for total tests on herds, total cattle slaughtered, new herd incidents, and herds not TB free, were analysed globally for GB and locally for the constituent regions of Wales, Scotland, West, North, and East England. After detecting trends over time, underlying regional differences were compared with the testing policies in the region. Total cattle slaughtered are decreasing in Wales, Scotland and West England, but increasing in the North and East English regions. New herd incidents, i.e. disease incidence, are decreasing in Wales, Scotland, West English region, but increasing in North and East English regions. Herds not TB free, i.e. disease prevalence, are increasing in West, North, and East English regions, while they are decreasing in Wales and Scotland. Total cattle slaughtered were positively correlated with total tests in the West, North, and East English regions, with high slopes of regression indicating that additional testing is likely to facilitate the eradication of the disease. There was no correlation between total cattle slaughtered and total tests on herds in Wales indicating that herds are tested frequent enough in order to detect all likely cases and so control TB. The main conclusion of the analysis conducted here is that more frequent testing is leading to lower TB infections in cattle both in terms of TB prevalence as well as TB incidence.

> *A man who practised the pentathlon, but whom his fellow citizens continually reproached for his lack of bravely, migrated to the overseas. After some time he returned, and went around boasting of having accomplished many extraordinary achievements in various cities, but above all having made such a jump when he was in Rhodes that not even an Olympic Champion could equal it. And he added that he could provide as witnesses of his achievement people who have actually seen It, if they ever came to the city. Then one of the bystanders spoke out: 'But if this is true there is no need for witnesses: **Here is your Rhodes, make the jump**'. The fable shows that as long as one can prove something by doing, speculation is superfluous.*
>
> *Aesop (Αἴσωπος, 620-560 BP). The maxim is also known in Greek as 'ιδού η Ρόδος, ιδού και το πήδημα' and in Latin as 'hic Rhodus, hic salta'.*

     Disease spread is a complex process involving several factors such as social structure of the individuals forming the population, biological and epidemiological characteristics of both the population and the disease, any economic and geographic factors associated with the mobility of the population as well as the spatial scale of that movement (Christakos et al. 2006). Therefore, understanding the spread of any disease is a highly interdisciplinary exercise as biological, social, geographic, economic, and medical factors may shape the way a disease moves through a population and options for its eventual eradication (Christakos et al. 2006; Christakos et al. 2014; Hethcote 2000). Historically, *bovine tuberculosis* (TB, also abbreviated as bTB) was almost eradicated from the GB, with a minimum in the number of cattle herds containing an individual that reacted positively to a test for TB being reached in the late 1970s (Krebs et al. 1997). Since this time the disease has been steadily increasing in both its prevalence in the cattle herd and its spread around the country (Abernethy et al. 2013; Gilbert et al. 2005).

     Bovine TB is a chronic disease in cattle (Thrusfield 2013). Although TB only rarely poses a threat to human health it causes a serious food security challenge to the agricultural industry in terms of dairy and meat production (Marsden & Morley 2014; Tomlinson 2013): Cattle which are detected or thought to have a high chance of being TB infected are slaughtered and thus the farmer loses the value of the animal and its output, although in GB the government pays farmers compensation for slaughtered animals based on their market value and so a large part of the economic cost is borne by the taxpayer (DEFRA 2014b; Krebs et al. 1997). In addition there are costs associated with testing animals for TB. When an animal in a herd tests positive for the disease, the entire herd is put under movement restrictions until all the remaining animals have tested repeatedly negative (DEFRA 2014b; Krebs et al. 1997). Until that point cattle cannot be moved to market, nor can animals be brought into the herd to replace slaughtered animals (DEFRA 2014b; Krebs et al. 1997). Further, TB infected cattle may facilitate the spread of infection to other domestic and wild animals (Caron et al. 2013; McCallan et al. 2014). Within the past 10 years over £0.5 billion on testing, compensation and research with further costs being borne by the agricultural industry (Godfray et al. 2013).

In order to control and eradicate TB from cattle, all cattle are regularly tested via a skin test at a frequency that is determined by the prevalence of TB within the local cattle population - the primary ante-mortem test for TB in the GB. In addition, all cattle moved to another farm are also tested prior to movement (Krebs et al. 1997). According to EU Directives 64/432/EEC and 97/12/EC the minimum testing frequency for cattle depends on the percentage of infected cattle herds: Annual testing is required unless the percentage of infected herds in a state or region of the state is 1% or less. When the percentage of infected herds is 0.2% or less than 0.1% testing may be conducted every three or four years respectively. In practice, most places in the GB tested cattle every four years during the 1990s (DEFRA 2014b; Krebs et al. 1997). Increasing test frequency would increase the annual cost of testing (Krebs et al. 1997). All cattle moving from one farm to another are tested prior to movement and the ones detected as infected are removed and slaughtered and the herd of origin is placed under movement restrictions (AHVLA 2013). The skin test on cattle is known to be imperfect - the test will not always detect an infected individual - and the presence of a common parasite *Fasciola hepatica* is reported to under-ascertain the rate of the skin test by about one-third (Claridge et al. 2012). Studies evaluating the sensitivity of the test suggest that its sensitivity lies between 52% and 100% with median values of 80% and 93.5% for standard and severe interpretation, respectively (DEFRA 2009). As a result many positive TB cases are first discovered in slaughterhouses.

Spatio-temporal disease analysis (Lin et al. 2015; Moustakas & Evans 2015; Ruiz-Medina et al. 2014) in terms of time trends and geographic disparities of disease dynamics may provide useful insights into the overall efficiency of control efforts and eradication of the disease (Booth et al. 2005) as well as the relative efficiency of different management measures towards eradication (Yu et al. 2013). In the case of TB monthly statistics are recorded as part of legal requirements of the EU Member States (DEFRA 2014b) and these data are publicly available. In October 2009 Scotland was officially declared as TB free while the rest of GB was not (Abernethy et al. 2013). Scotland has had a risk based surveillance testing policy (Dorjee et al. 2013) under which high risk herds are tested frequently while lower risk herds are tested every four years and since September 2009 has been designated as TB free . Wales has had an annual or even more frequent testing policy of all herds since January 2010 while in England the number of herds tested annually (instead of every four years) has increased since 2010 and expanded further in the East and North regions in 2011; however several areas are still tested every four years (DEFRA 2014b). Thus, regional differences in TB detections may provide insights of different policies against eradicating the disease. In the current study we sought to quantify seasonal trends of TB in cattle population across different regions of GB and link potential differences to the control strategies operated in the three countries and regions of England.

**Methods**

*Data*

Publicly available time series of monthly bovine TB statistics in cattle in GB published by DEFRA were used. The bovine TB statistics are presented for GB by different geographical areas with regional-level statistics provided for Scotland, Wales, and North, East and West English region (DEFRA 2014a). Statistics for GB (without distinguishing between regions) are available from January 1996 to June 2014, while the regional statistics are available from January 2008 to June 2014, all on a monthly time step (DEFRA 2014a). Statistics are affected by seasonal patterns and variations in the frequency of testing. TB testing is seasonal as more herds are tested in the winter when more cattle are housed inside. The animals tested are not a random sample of the whole GB herd: Herds are tested more frequently in areas of higher TB incidence than in those with a historically low incidence and thus herds experiencing several separate incidents in the same reporting period will appear more often in the dataset (see below 'Herds not TB free'). In general, as more tests are carried out, more TB infected herds are likely to be found (DEFRA 2014a). From the available datasets the following monthly statistics were analysed:
(1) *Herds not TB free*: The number of herds that did not have TB free status (i.e. were under movement restrictions) at some time during the period shown due to a TB incident. This figure may include herds experiencing a TB incident that began before (but continued into) the report period. Likewise, any herds experiencing two separate incidents in the same reporting period could appear twice under this data column.
(2) *Total cattle tests*: The count of the number of cattle for which a tuberculin skin test has been carried out. An individual animal could be tested more than once in each time period. All test types are included except for slaughterhouse cases.
(3) *Total tests on herds*: Herds for which tuberculin skin testing is carried out on at least one animal in a herd during the period listed. The following test types are not included in this measure: Pre- and post-movement, gamma interferon blood tests, and private tests.
(4) *New herd incidents*. The disclosure of at least one test reactor or a confirmed slaughterhouse case in a herd that was previously TB free during the period shown.
(5) *Total cattle slaughtered*: This is the number of infected cattle slaughtered and it is the sum of reactors slaughtered + inconclusive reactors slaughtered + direct contacts slaughtered. *Reactors slaughtered* include cattle which were compulsorily slaughtered by AHVLA because it responded to a relevant test for TB in a way that was consistent with it being TB infected. In any given period, the majority of animals are slaughtered as reactors to the tuberculin skin test. A much smaller number of reactors are those giving a positive result to the ancillary interferon-gamma blood test. *Inconclusive reactors slaughtered* are cattle showing a positive reaction to TB that was not strong enough for it to be deemed a reactor. However, instead of being tested again after 60 days (the normal procedure in this case) the animal was voluntarily slaughtered by the owner. *Direct contacts slaughtered* are cattle from an officially TB free status withdrawn herd that, although not a test reactor, was considered to have been exposed to TB and compulsorily slaughtered.

Details of the methodologies and revisions of the methodologies through time can be found at (DEFRA 2014a) and references therein. All data are freely available at: https://www.gov.uk/government/statistics/incidence-of-tuberculosis-tb-in-cattle-in-great-britain.

*Time series decomposition*

We analysed the five monthly variables (*Total cattle tests, Total tests on herds, Total cattle slaughtered, Herds not TB free*, and *New herd incidents*) using decomposition time series analysis (Ye et al. 2015). In order to disentangle the actual variance of the signal due to stochastic fluctuations in the time series data (Koutsoyiannis 2011) across time scales (Markonis & Koutsoyiannis 2013) qualitative diagnostics were used to make a deterministic attribution of linear trends (Koutsoyiannis 2005). In order to detect linear trends in time series data we estimate the variance change points of the time series using the method of Lavielle (1999, 2005). In brief the method identifies breaking points in the data such that linear models can be fit within these points (Lavielle 1999, 2005). The method assumes that the data follow a normal distribution and detects simultaneous changes in the mean and variance of the distribution but requires no assumptions about the dependence structure of the time series (i.e. linear, or non-linear trends). After the detection of breaks, the time series is split into a seasonal and trend component (Lavielle 1999, 2005). The method of Lavielle (1999) finds the best segmentation of a time series, given that it is built by *K* segments. It searches the segmentation for which a contrast function (measuring the contrast between the actual series and the segmented series) is minimized. The contrast function used for measuring different aspects of the variation of the series from one segment to the next was calculated accounting for both the mean and the variance varying between segments. It is required to specify a value for the minimum number of observations *Lmin* in a segment (here *Lmin* = 12 months), as well as the maximum number of segments *Kmax* in the series. In order to estimate the best number of segments *Kmax* to partition the time series the decrease of the contrast function with the number of segments is examined. An automated way to estimate the optimal number of segments relying on the presence of a 'break' in the decrease of the contrast function has been suggested (Lavielle 2005). According to this method the last value of *K* for which the second derivative of a standardized contrast function is greater than a threshold *S* is chosen - see (Lavielle 2005) for details. Based on numerical experiments, it has been proposed to choose the value *S* = 0.75 (Lavielle 2005). As for some short time series it has been indicated that this value may not be optimal and may depend on the value of *Kmax* (Lavielle 2005), we also performed graphical examination of the decrease of the contrast function with the number of segments. See e.g. Sur et al. (2014) for an application of the method in ecology, and Lijffijt et al. (2014) for segmentation of time series.

Having determined segments (time periods within the data) in which linear trends exist, seasonal filters were used to de-seasonalize a time series using a multiplicative decomposition (Brockwell & Davis 2002) . Decomposition in time series analysis is used in order to separate the time series into linear trend and seasonal components, as well as error & stochasticity. As the seasonal pattern in the data depended on the level of the data - more cattle tests are carried out during winter months and in general as more tests are carried out, more TB infected herds are likely to be found (DEFRA 2014a) – a multiplicative model structure was employed accounting for this effect. A multiplicative de-trending model is used when the size of the seasonal pattern in the data depends on the level of the data (Carstensen 2005; Gocheva-Ilieva et al. 2014). The model assumes that as the data increase, so does the seasonal pattern. Several time series exhibit such a pattern (Barucci et al. 2010). The multiplicative model is expressed as: $Y_t$ = *Trend * Seasonal * Error*, where $Y_t$ is the observation at time *t*. An approximate confidence interval (95%) of the prediction using a rule of thumb *prediction ± 2·x Standard Error* were also plotted. All data were $\log_{10}(x+1)$ transformed prior to the analysis and successfully tested for normality.

In order to assess model fit to the data the Mean Absolute Percentage Error (MAPE) index was used. MAPE measures the accuracy as a percentage of the error (Armstrong & Overton 1977) and is expressed as: MAPE = $\frac{\sum |y_t - \hat{y}_t|}{n} \times 100, (y_t \neq 0)$, where $y_t$ is the actual value and $\hat{y}_t$ the fitted value. Because this number is a percentage, it may be easier to understand than other performance measures. For example, if the MAPE is 5, on average the prediction is out by 5%.

*Analysis between total tests on herds and total cattle slaughtered*

In addition we sought to investigate the relation between the number of total tests on herds or total cattle tests and total cattle slaughtered. We sought to quantify whether there are geographic regions in GB where a higher number of tests does not lead to the detection of more infected herds/cattle. While in general as more tests are carried out, more TB infected herds are likely to be found (DEFRA 2014a), an absence of correlation between number of tests and number of detections can provide statistical inference that herds are tested sufficiently often so that additional testing would not facilitate the detection of TB infected herds/cattle and a higher slope in the regression implies that more frequent testing would lead to higher detection of TB infected herds/cattle (García-Portugués et al. 2014; Guo et al. 2006; Weisberg 2014). We therefore examined the slopes of linear regressions between total cattle slaughtered vs. total tests on herds and total cattle tests. In addition a local regression line was fitted between total cattle slaughtered and total tests on herds. A local regression helps explore the potential relationships between two variables without fitting a specific model, such as a regression line or a theoretical distribution and thus deviations from linearities could be explored, as well as the effect of extreme values or outliers in the regression analysis (Abaurrea et al. 2015; Cleveland 1979). Local regressions performed here are calculated using the lowess smoothing method. Lowess (locally-weighted scatterplot smoother) works by selecting a fraction (here f = 0.4 & 0.6, and number of steps = 1 & 3 respectively) of all points, using the data closest in x-value on either side of the (x,y) point. For each data point a weighted linear regression was performed, giving points closest to each x-value the most weight in the smoothing and limiting the effect of outliers.

**Results**

*Trends in the raw data between and within years*

Mean values and 95% confidence intervals of the mean of the raw data (i.e. not de-trended and log-transformed data) are shown in Fig. 1 (values between years) and Fig. 2 (values per month). In terms of between years values, total tests on herds are increasing on an annual basis in all regions (Wales, West, North, and East of England), except Scotland where the total tests on herds are decreasing (Fig. 1a). Total cattle slaughtered are increasing in the North and in the past three years in the East English region (Fig. 1b). There are no consistently clear patterns in total cattle slaughtered in Scotland, West England, and Wales (Fig. 1b). Herds not TB free are increasing in the North and East English regions and oscillate in Wales, Scotland and West English region (Fig. 1c). New herd incidents are increasing in North England while no clear trend in mean values is recorded in any other region (Fig. 1d). Seasonal (within the year) patterns of total tests on herds and total cattle tests have similar patterns in all GB regions: Testing is peaking during winter months when winter housing is applied, declining until August and gradually increasing thereafter (Fig. 2a). Total cattle slaughtered does not follow this pattern as in most cases mean values do not exhibit a clear trend prior to de-trending (Fig. 2b). Herds not TB free in terms of mean values are peaking in late winter months (March, April) and declining thereafter across all regions (Fig. 2c). New herd incidents are consistently higher during winter months across all regions (Fig. 2d).

*Patterns following de-trending time series analysis*

*Number of cattle slaughtered*

Once the raw data have been de-trended it can be seen that total cattle slaughtered in GB are increasing; the rate of increase was higher during 1996 – 2001 (variance break point September 2001, MAPE = 0.89) than in from Sept 2001 – June 2014 (MAPE = 2.42); (Fig. 3a). Within the year variance as indicated from seasonal components was also larger during 1996-2001 (Fig. 3a). No break points were detected for the regional trends (all could be approximated with a single linear trend). Total cattle slaughtered are decreasing in Wales (MAPE = 2.92), Scotland (MAPE = 0.36) and West England (MAPE = 3.17); (Fig 3b-d), but increasing in the North (MAPE = 3.62) and East (MAPE = 3.82) English regions (Fig. 3 e and f). There are some cases where model residuals are autocorrelated as for example in the decomposition of total cattle slaughtered Scotland. As for the decomposition of regional data $\log_{10}(x+1)$ transformation produced good model fit for most regions we decided against applying more stringent data transformations for the cases where autocorrelated residuals exist and proceed with a consistent analysis for all regions and data. Residuals of the decomposition are provided in Supp. 1 'Residuals of Total cattle slaughtered'.

*New herd incidents – TB incidence*

New Herd incidents are increasing in GB throughout from 1996 – 2014, and the within year variance has been lower during January 1996 to October 2001 (variance break point, MAPE = 0.79) and increasing from November 2001 to June 2014 (MAPE = 1.34); (Fig. 4a). No break points were detected for the regional trends (all could be approximated with a single linear trend). New herd incidents, and thus disease incidence, are decreasing in Wales (MAPE = 0.89), Scotland (MAPE =

0.91); (Fig 4 b and d), and marginally decreasing in West English region (MAPE = 1.86); (Fig 4 c), but increasing in North (MAPE = 1.42) and East (MAPE = 2.89) English regions (Fig. 4 e and f). Residuals of the decomposition are provided in Supp. 1 'Residuals of New herd incidents'.

*Herds not TB free – TB prevalence*

Herds not TB free were increasing in GB from January 1996 – January 2002 (break point, MAPE = 0.61), followed by a steeper increase during February 2002 – December 2006 (break point, MAPE = 2.41), and from January 2007 – June 2014 (MAPE = 1.07) are also increasing but at a lower rate previously (Fig. 5a). During the last period of Jan 2007 – June 2014 the within the year variance is also lower (Fig. 5a). No break points were detected for the regional trends (all could be approximated with a single linear trend). Herds not TB free, and thus disease prevalence, are increasing in West (MAPE = 1.98), North (MAPE = 0.99), and East (MAPE = 1.81) English regions (Fig 5 c, e and f), while they are decreasing in Wales (MAPE = 0.41) and Scotland (MAPE = 2.82); (Fig. 5 b and d). In general there was autocorrelation in the residuals of the decomposition of herds not TB free in most regions. Residuals of the decomposition are provided in Supp. 1 'Residuals of Herds not TB free'.

*The relationship between total tests on herds and total cattle slaughtered*

Total cattle slaughtered were positively correlated with total tests conducted on herds in the North and East English regions, with high slopes of regression (Fig. 6). Total cattle slaughtered were also positively correlated with tests in Scotland and in West England, although in West England the slope was considerably lower (Fig. 6). There was no correlation between total cattle slaughtered and total tests on herds in Wales (Fig. 6). For further details regarding regression analysis between total cattle slaughtered and total tests on herds per region and figures at a higher resolution please refer to Supp. 1 'Regression analysis'. These results are also held when correlating the total cattle tests against the total cattle slaughtered (results not shown here).

**Discussion**

There are major spatial differences between the temporal trends of the dynamics of TB in cattle in the GB and thus spatio-temporal analysis provides useful insights into linking those differences with the underlying control strategies (Christakos & Hristopulos 1998; Christakos et al. 2014; Yu et al. 2013). In order to account for both TB prevalence and incidence in cattle we have analysed both herds not TB free, and new herd incidents. Incidence is rate of occurrence of new cases, and so new herd incidents were used to asses this. Prevalence is the proportion of the total number of cases to the total population and is more a measure of the burden of the disease on society with no regard to time, and herds not TB free (restricted) were used as a proxy of prevalence.

The majority of TB infected herds and cattle in GB are located in England (West, East, and North English regions), while both TB incidence and prevalence are decreasing in Wales and Scotland. In the West English region new herd incidents are marginally decreasing but herds not TB free are increasing indicating that once a herd is infected, the TB infection remains in the herd. The overall increase pattern of both TB incidence and prevalence in cattle in GB is thus driven by the English regions. Scotland and Wales both have a declining number of new herd incidents as well as herds not TB free and thus the current programme applied, all else being equal, appears to be leading to eradication or control of the disease. In Wales in particular for the period of January 2008 to December 2009, prior to applying annual or more frequent testing, the mean of new herd incidents was 99.33 and the mean of herds not TB free was 979.41, while for the period of January 2010 to June 2014 when annual or more frequent testing was applied the mean of new herd incidents was 83.57 and the mean of herds not TB free 863.31 (Fig. 7). All regional data analysed here could be approximated with a single linear trend, while the same data for GB were not; linear trends were fitted for total cattle slaughtered and new herd incidents (break points in Sept 2001, and Nov 2001 respectively), while herds not TB free had two break points in Jan 2002 and Dec 2006. The first break points are temporally close to the TB testing interruption due to the foot-and-mouth disease outbreak in GB in 2001. Both TB incidence and prevalence was affected by interrupted cattle testing: new herd incidents have a larger within year variance after the break point of Nov 2001, while herds not TB free had the highest rate of increase during Jan 2002 – Dec 2006. This may serve as an inverse proof of the effects of testing on TB dynamics in cattle because when testing was removed TB infections rose so substantially.

Scotland applies a risk based surveillance policy (Oidtmann et al. 2013) testing more often at high risk areas (DEFRA 2014b). This policy appears to be facilitating TB eradications as there are diminishing new herd incidents as well as number of herds not TB free. In addition Scotland is the only region in the GB where the total number of tests on herds has been decreasing through the period under consideration, while in all other regions it is increasing either due to the increasing number of new herd incidents (North and West English regions) and/or due to the introduction of an annual testing policy (Wales). Thus the risk based testing policy not only is leading to TB eradication but it is also cost effective as decreased numbers of TB infected herds are achieved via a lower total number of tests (Bessell et al. 2013). However, in Scotland the total number of cattle per hectare is low (Yang et al. 2014) and it remains unknown whether such testing policy would lead to TB eradication when the total number of cattle and subsequently the total cattle tests is high.

Cases in human epidemiology have concluded that in all but the lowest-risk populations, routine screening is justified on both clinical and cost-effectiveness grounds (Paltiel et al. 2005). In addition, risk based versus routine based surveillance in human epidemiology has indicated that routine based testing provides a better alternative (Beckwith et al. 2005; Jenkins et al. 2006). While there certainly are fundamental differences between human and animal epidemiology due to behavioural and social structure differences, a positive albeit weak correlation between total cattle slaughtered and total tests on herds/cattle was recorded in Scotland. This provides statistical inference for a potential for higher detection of TB infected herds/cattle if more tests were employed (Guo et al. 2006; Weisberg 2014).

Both the number of new herd incidents and number of herds not TB free are declining in Wales and the number of the total cattle slaughtered is not correlated to the total tests on herds or total cattle tested. Thus, more cattle tests would not have resulted in detecting more infected herds/cattle (Weisberg 2014). Interestingly, Wales is the only region in the GB that has adopted annual testing policy in all herds since Jan 2010, and in some counties in Wales cattle testing is exercised even every six months (DEFRA 2014b). Indeed there is a significant difference between pre and post Jan 2010, when at least annual testing was applied, of both new herd incidents and herds not TB free in Wales (Fig. 7).

*Synthesis and management implications*

1. *Testing frequency*

The main result derived here from statistical analysis of publicly available data from the British Government show that increased cattle testing leads to TB control or possibly eradication as exemplified by the results in Wales. This conclusion fully supports the outputs of a computational model suggesting that all eradication scenarios included cattle testing frequencies of annual or even more frequent testing (Moustakas & Evans 2015). This is the cattle testing frequency exercised in Wales since 2010, with annual or even more frequent testing applied (DEFRA 2014b).

2. *Within year effects & winter housing*

The pattern of new herd incidents is clearly peaking during winter months (Fig. 2d and de-seasonalised effects in Fig. 4). Herds not TB free are also peaking during late winter months (Fig. 2c and de-seasonalised effects in Fig. 5). However total tests on herds are also peaking during winter months (Fig. 2a). The fact that new herd incidents peak during winter months may support three hypotheses that merit further investigation: (i) TB prevalence in cattle is increased during winter months when winter housing is practised - i.e. all cattle within a farm are summoned in a winter housing facility with high densities of cattle within a small surface area. This has been suggested to increase cattle-to-cattle TB spread from Moustakas & Evans (2015). (ii) Herds not TB free and new herd incidents are peaking late in the winter simply because more tests are conducted during these months. Then this implies that the more one tests, the more infected cattle is likely to detect and thus should test more often. (iii) Lower testing during summer months when cattle are out in the field may create buffers of infection between cattle; cattle that get infected during summer months will more likely be detected during winter months as more tests are conducted then. During that interval they may infect other cattle as they will not be removed from the herd and the herd will not be under movement restrictions. Interestingly new herd incidents as well as total tests on herds are lowest during summer months, when cattle are out in the field, the period that interactions with badgers are maximised. If badgers are the agent of cattle infection it is logical to test cattle during summer months.

The issue of winter housing while it was only recently proposed by (Moustakas & Evans 2015) in animal TB epidemiology is not new in human TB epidemiology. Indeed several studies have found seasonality in TB and attributed it mainly to indoor winter crowding and lower vitamin D levels with lower levels of sunshine (Koh et al. 2013; Parrinello et al. 2012; Wingfield et al. 2014). There are also cases where winter crowding was ruled out (Top et al. 2013) but in any case this merits serious investigation: winter crowding was known to be a main factor of seasonal pick in infections and the fact that it was found not to be significant was a publishable result (Top et al. 2013). In order to quantify seasonal effects in cattle infections the analysis can include social networks of cattle (Corner et al. 2003; Croft et al. 2011) during winter housing and during summer months.

3. *The importance of public data*

We would like to highlight the importance of pubic data. In order to predict and mitigate disease spread informed decisions are needed. These decisions need to be taken based on data analysis and predictive models calibrated with data (Evans et al. 2014; Evans et al. 2013; Lonergan 2014). Our view is that making regional TB data available so that potential differences and underlying management decisions is a very good step forward. We argue that making publicly available the data regarding badger culling experiments as agents of TB infecting cattle will greatly facilitate their analysis and to informed decisions regarding TB control in GB.

**Acknowledgements**

We thank Yannis Daliakopoulos and Manolis Grillakis for their help during time series decomposition. Their code was developed under the auspices of the CASCADE Project (Tsanis & Daliakopoulos 2015). Comments from two anonymous reviewers considerably improved an earlier manuscript draft.

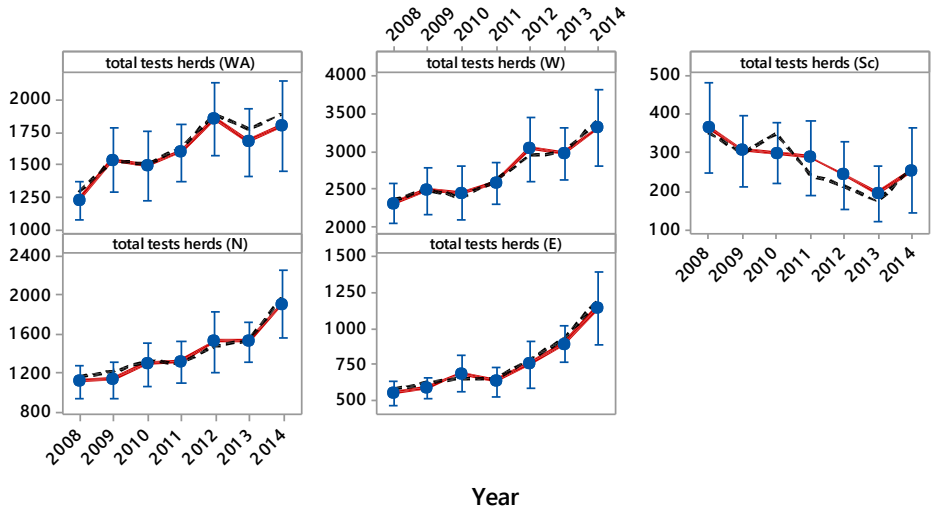

a. Total tests on herds - between years
95% CI for the Mean

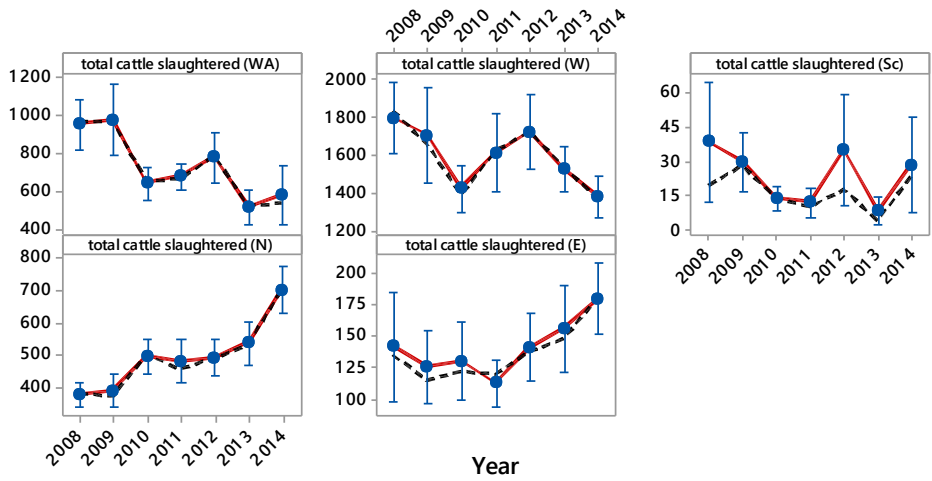

b. Total cattle slaughtered - between years
95% CI for the Mean

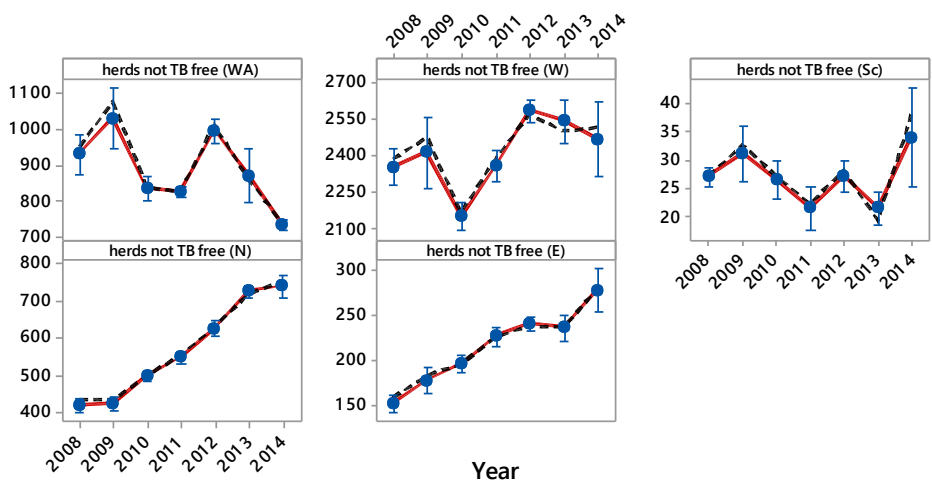

c. Herds not TB free - between years
95% CI for the Mean

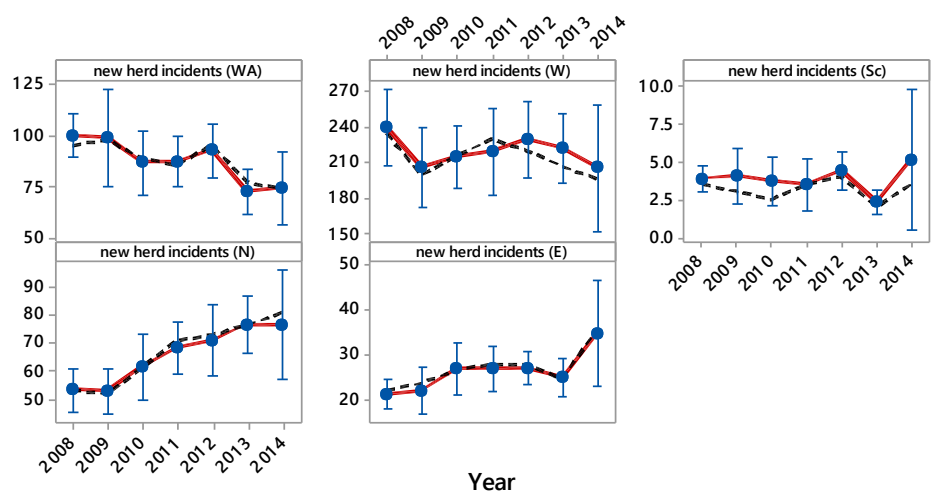

d. New herd incidents - between years
95% CI for the Mean

**Figure 1.** Confidence intervals of 95% of the mean values of **(a)** total tests on hers, **(b)** total cattle slaughtered, **(c)** herds not TB free, and **(d)** new herd incidents per year. Data per region in the GB from Jan 2008 to June 2014 recorded on a monthly basis. Wa: Wales, W: West English Region, Sc: Scotland, N: North English Region, E: East English Region. Intervals show the 95% confidence envelope of the mean, solid red lines connect mean values, and grey dotted lines connect median values between years.

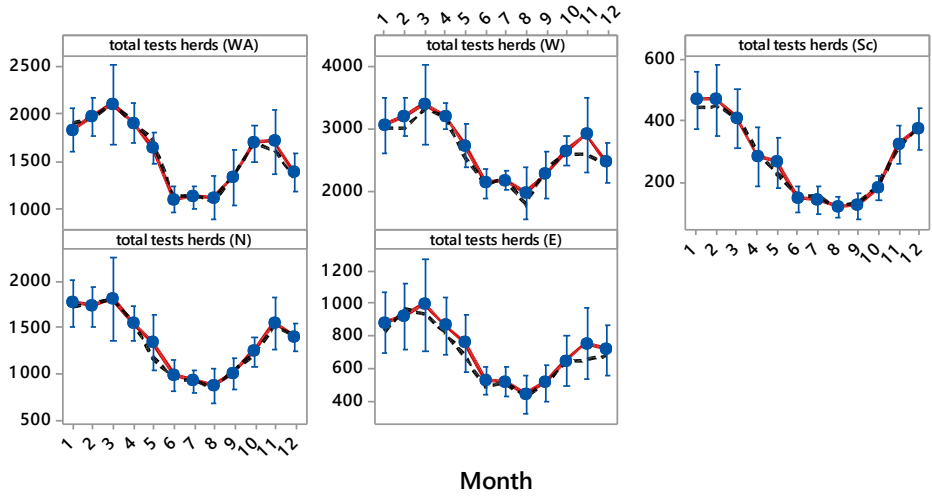

a. Total tests on herds - within year
95% CI for the Mean

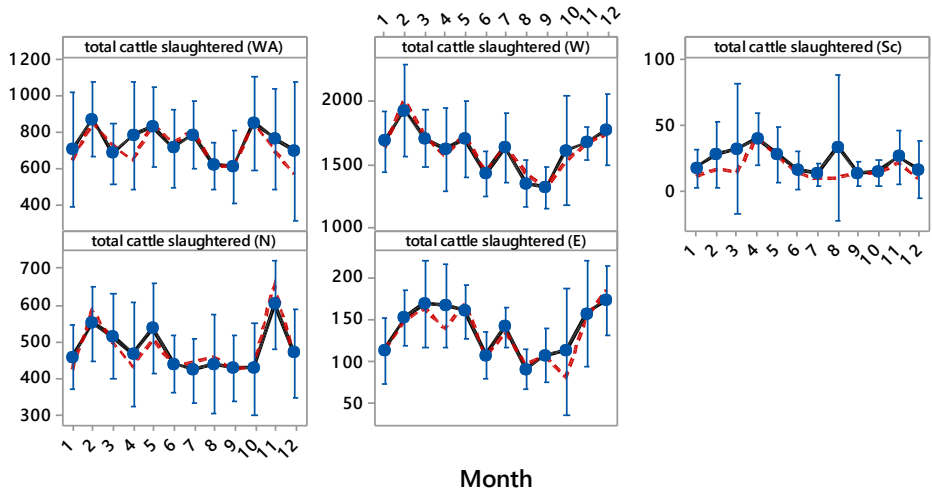

b. Total cattle slaughtered - within year
95% CI for the Mean

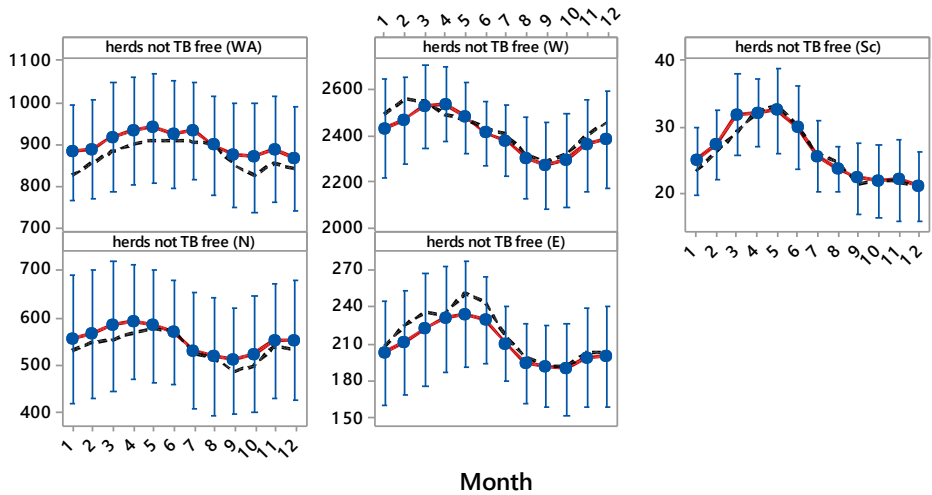

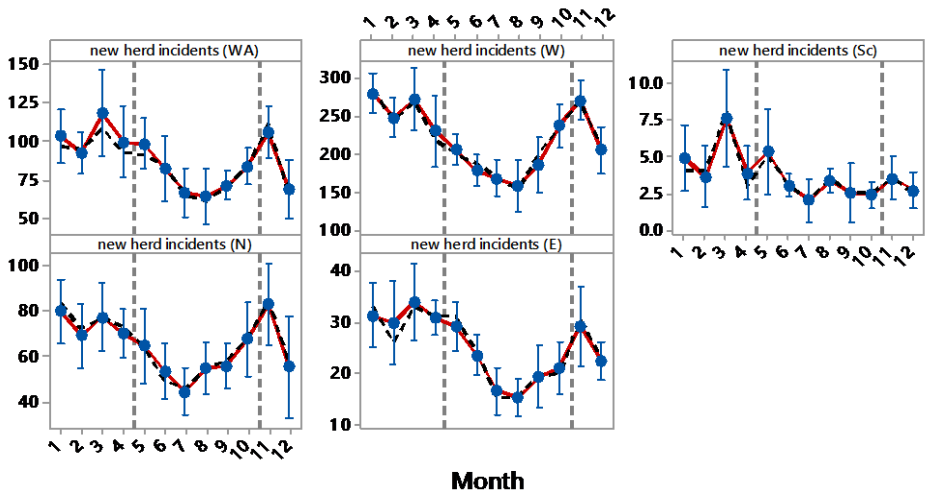

**Figure 2.** Confidence intervals of 95% of the mean values of **(a)** total tests on hers, **(b)** total cattle slaughtered, **(c)** herds not TB free, and **(d)** new herd incidents per month. Data per region in the GB from Jan 2008 to June 2014 recorded on a monthly basis. Wa: Wales, W: West English Region, Sc: Scotland, N: North English Region, E: East English Region. Intervals show the 95% confidence envelope of the mean, solid red lines connect mean values, and grey dotted lines connect median values between months. Dotted vertical lines on graph d (new herd incidents per month) show winter months (January – April & November - December) and summer months (May – October).

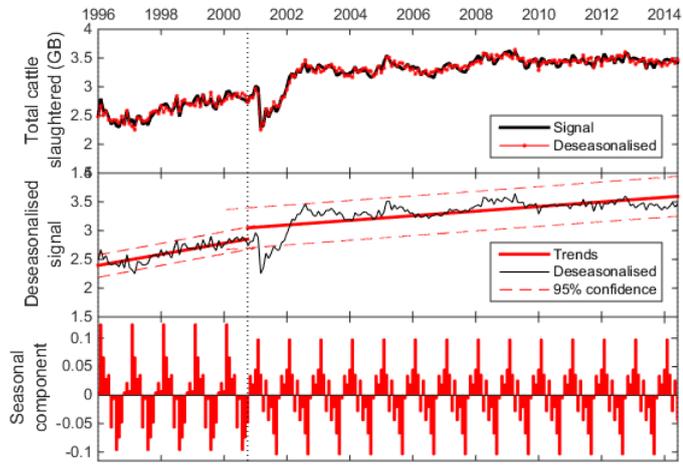

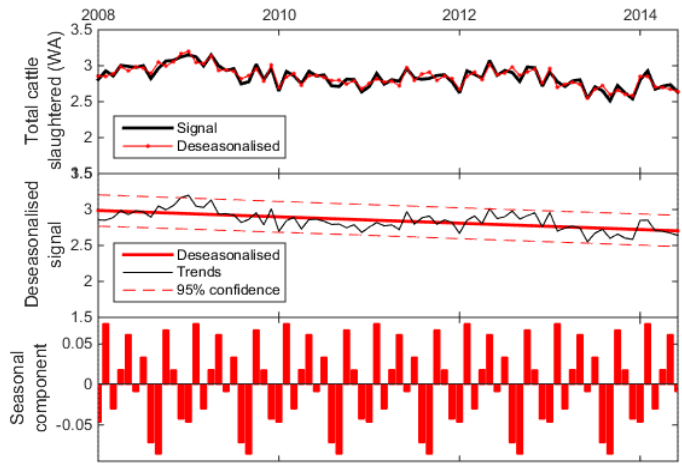

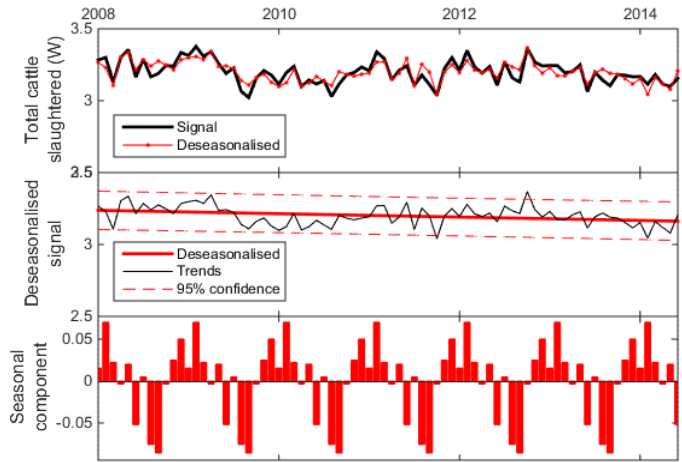

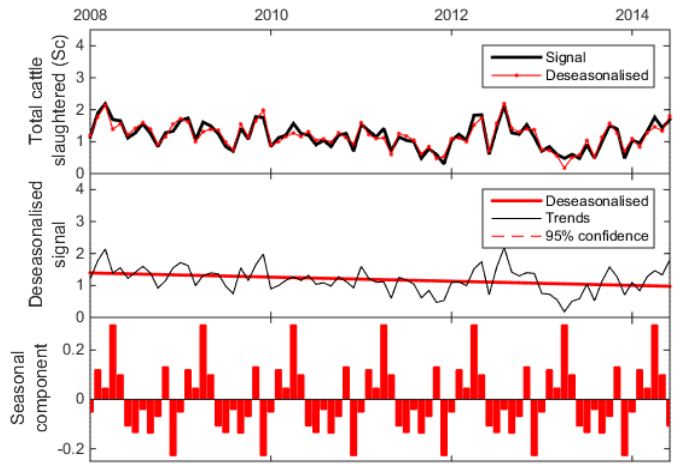

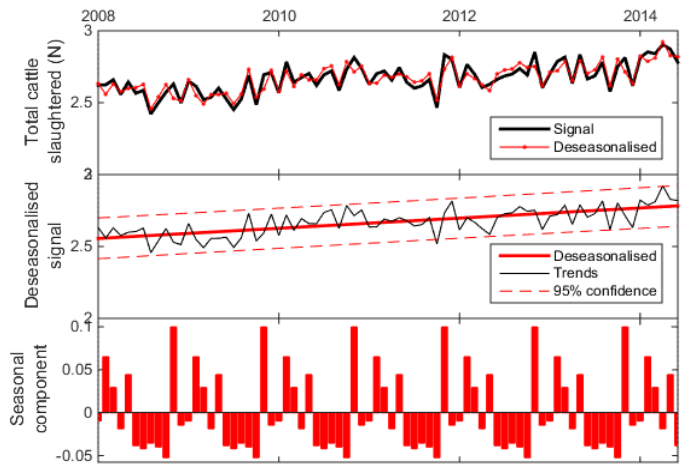

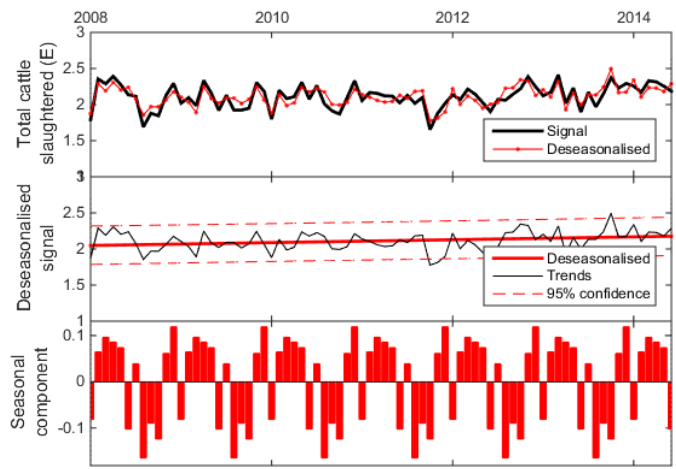

**Figure 3.** Time series decomposition of total cattle slaughtered (log-transformed data) in GB (data from Jan 1996 to June 2014) and in different regions in GB (data from Jan 2008 to June 2014) on a monthly time step. Upper panels of each graph show the $\log_{10}(x+1)$-transformed real data in black lines (*Signal*) and the model fit in red lines (*Deseasonalised*). Middle panels show a linear trend on the model fit with 95% confidence intervals of the trend. Lower panels show the seasonal (within the year) decomposition. Where applicable, vertical dotted lines indicate variance change points in the data and different linear trends were fitted for each period – see methods for details. Wa: Wales, W: West English Region, Sc: Scotland, N: North English Region, E: East English Region.

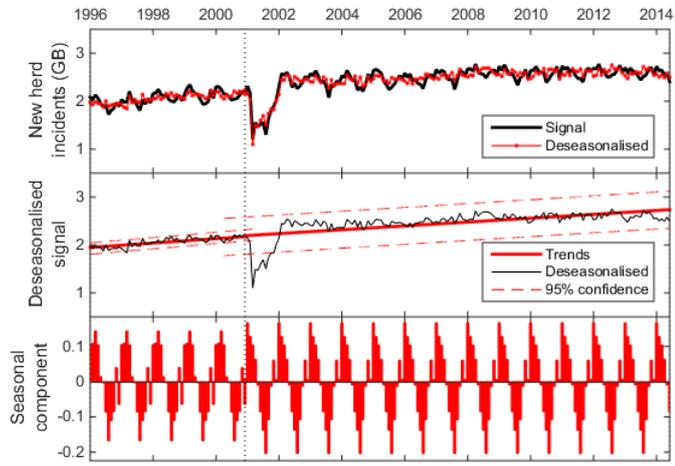

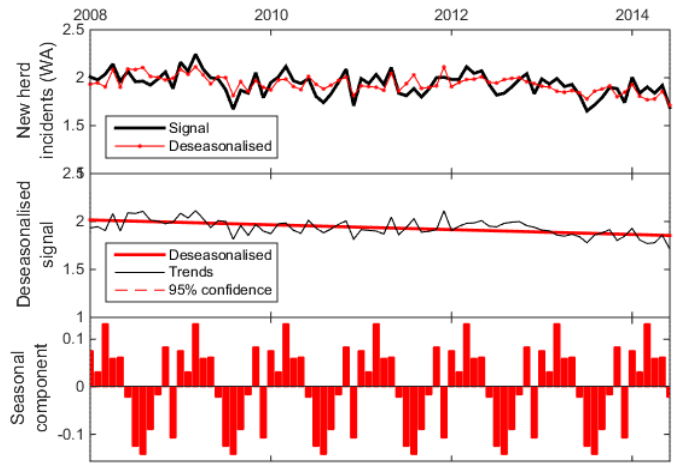

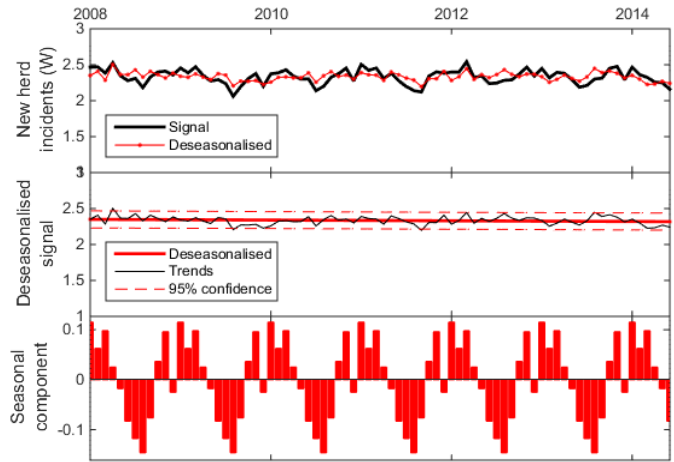

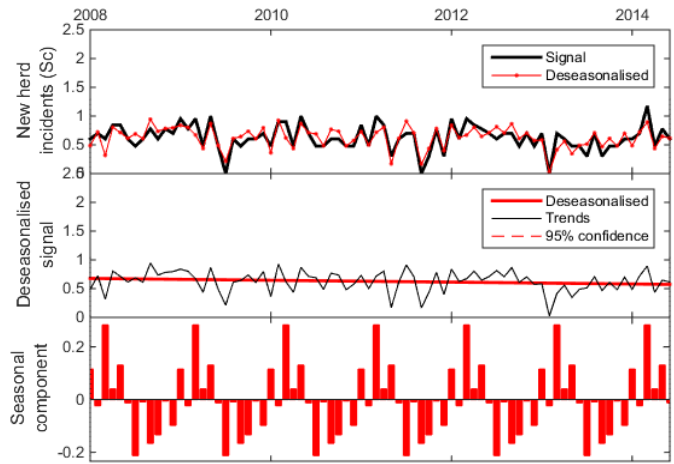

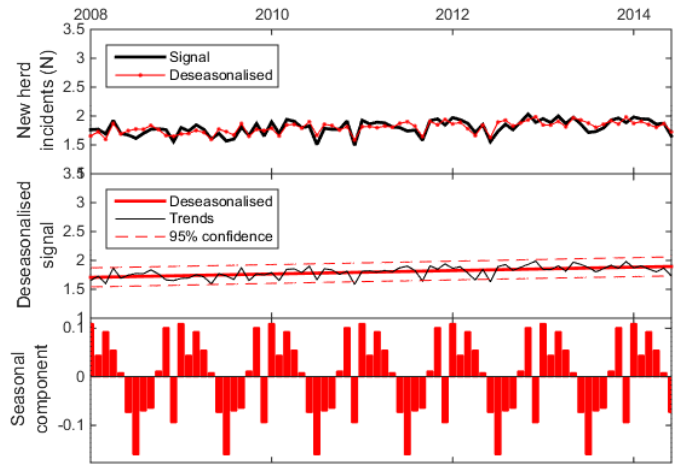

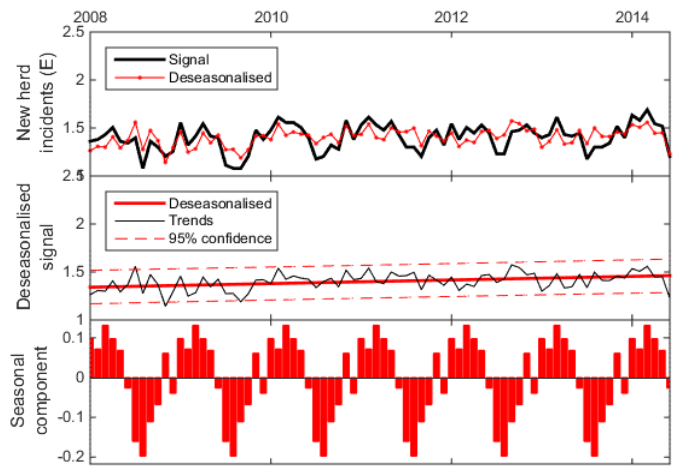

**Figure 4.** Time series decomposition of new herd incidents (log-transformed data) in GB (data from Jan 1996 to June 2014) and in different regions in GB (data from Jan 2008 to June 2014) on a monthly time step. Upper panels of each graph show the $\log_{10}(x+1)$-transformed real data in black lines (*Signal*) and the model fit in red lines (*Deseasonalised*). Middle panels show a linear trend on the model fit with 95% confidence intervals of the trend. Lower panels show the seasonal (within the year) decomposition. Where applicable, vertical dotted lines indicate variance change points in the data and different linear trends were fitted for each period – see methods for details. Wa: Wales, W: West English Region, Sc: Scotland, N: North English Region, E: East English Region.

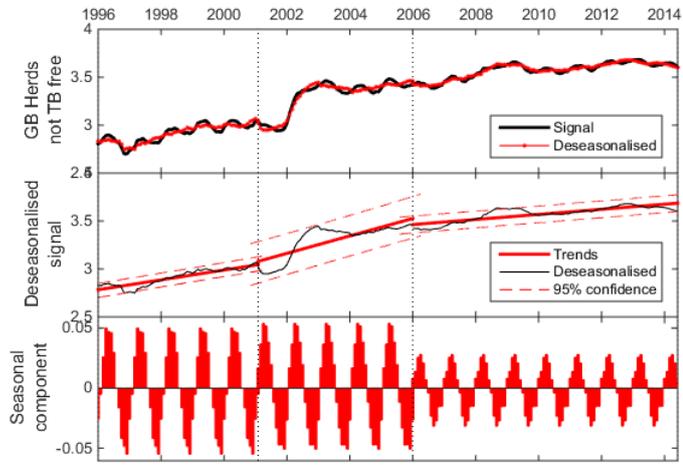

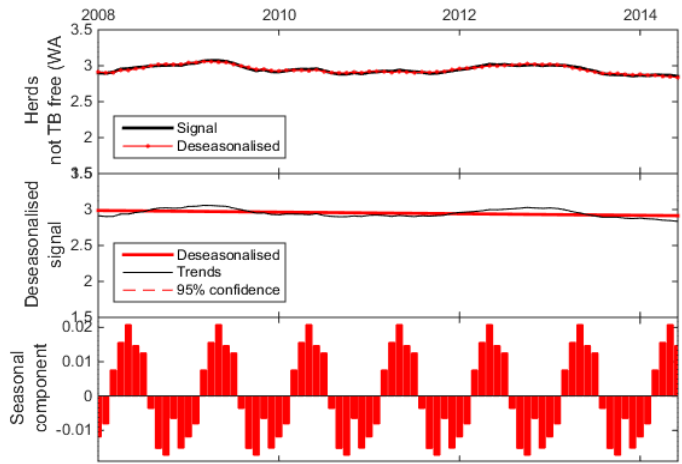

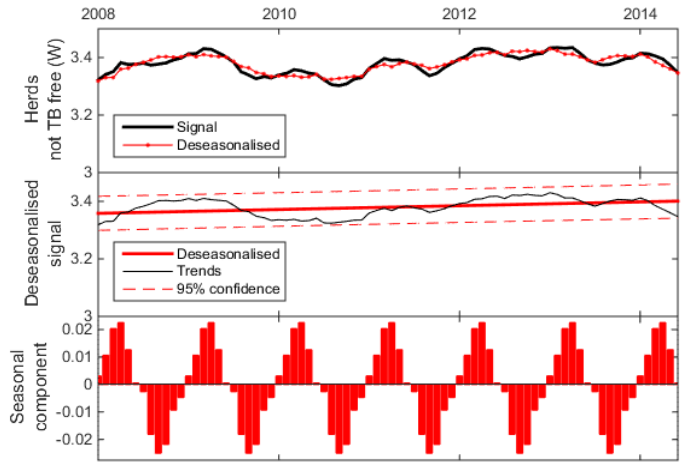

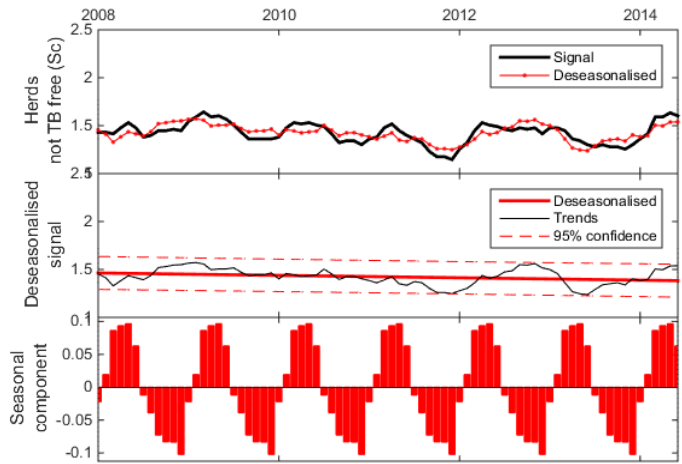

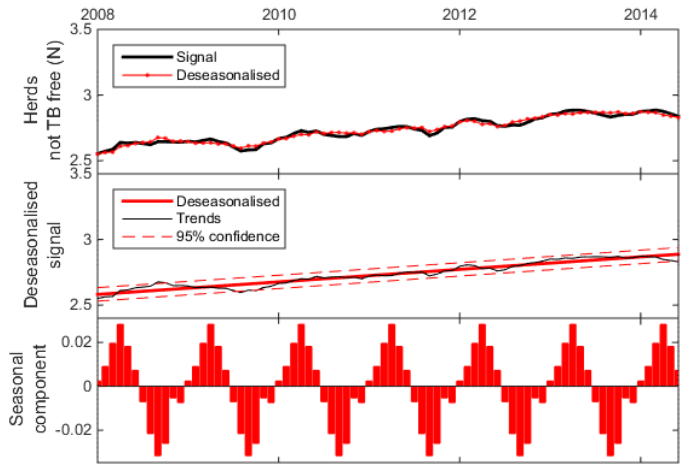

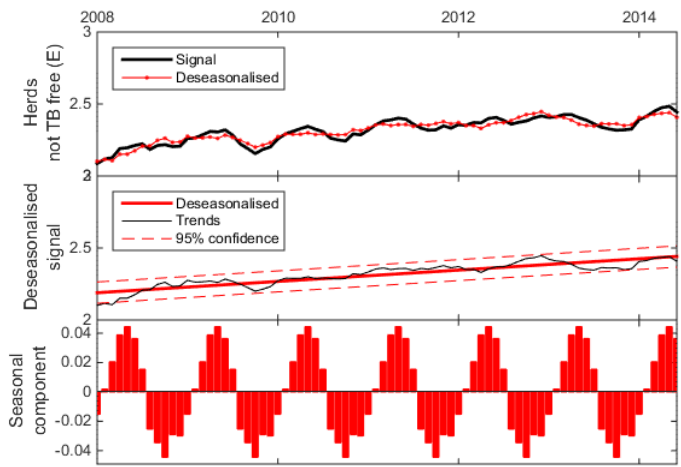

**Figure 5.** Time series decomposition of herds not TB free (log-transformed data) in GB (data from Jan 1996 to June 2014) and in different regions in GB (data from Jan 2008 to June 2014) on a monthly time step. Upper panels of each graph show the $\log_{10}(x+1)$-transformed real data in black lines (*Signal*) and the model fit in red lines (*Deseasonalised*). Middle panels show a linear trend on the model fit with 95% confidence intervals of the trend. Lower panels show the seasonal (within the year) decomposition. Where applicable, vertical dotted lines indicate variance change points in the data and different linear trends were fitted for each period – see methods for details. Wa: Wales, W: West English Region, Sc: Scotland, N: North English Region, E: East English Region.

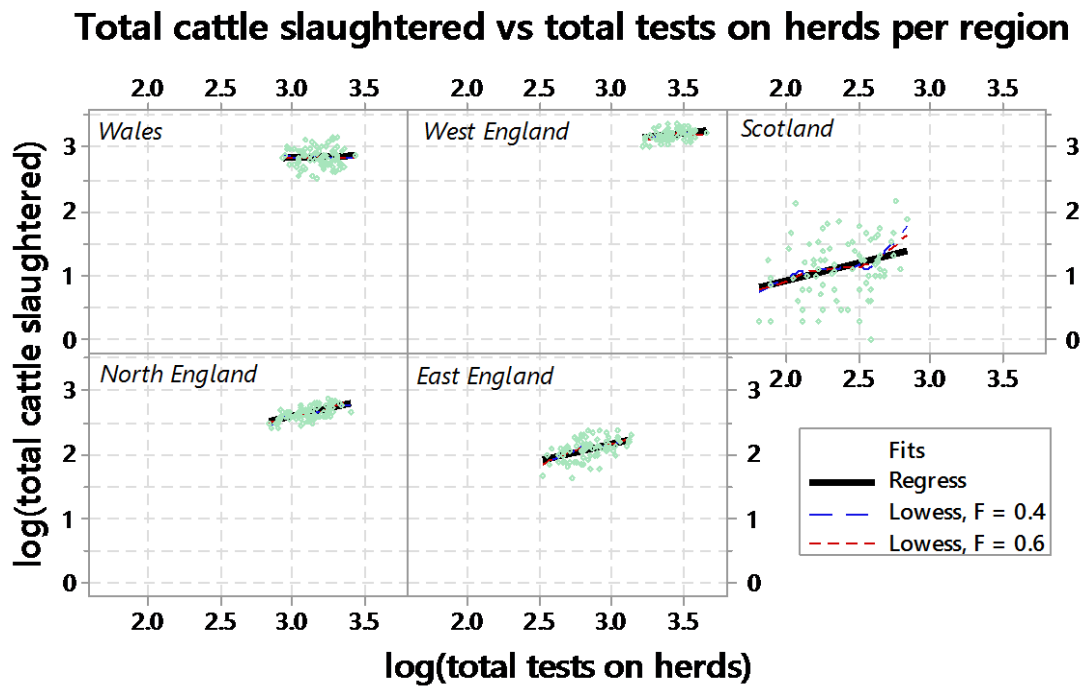

**Figure 6.** Linear regressions and locally-weighted smoother lines between total cattle slaughtered on the vertical axis and total tests on herds on the horizontal axis between regions in the GB. All panels are on the same scale. Regressions are plotted within the range of the data for each region. All data were $\log_{10}(x+1)$-transformed. For a higher resolution of each regression fit including confidence intervals please see 'Regression Analysis' in Supp. 1.

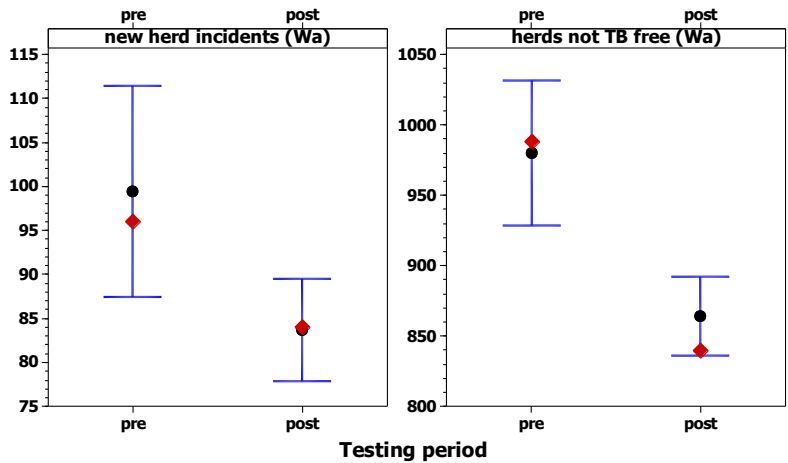

**Figure 7.** Confidence intervals of 95% for the mean value (in black circles) of New herd incidents (left panel) and Herds not TB free (right panel) in Wales. The median values are also listed in red squares. Pre and Post testing periods refer at the period before annual testing was applied in Wales (*Pre*: Jan 2008 – Dec 2009) and after introducing at least annual testing (*Post*: Jan 2010 – June 2014). The original data are plotted here (no log-transformations were applied).

**Supplement 1.**

**Regional and temporal characteristics of *Bovine Tuberculosis* of cattle in Great Britain.**

**By Aristides Moustakas & Matthew R. Evans**

**Residual plots of time series decomposition**

**Residuals of total cattle slaughtered**

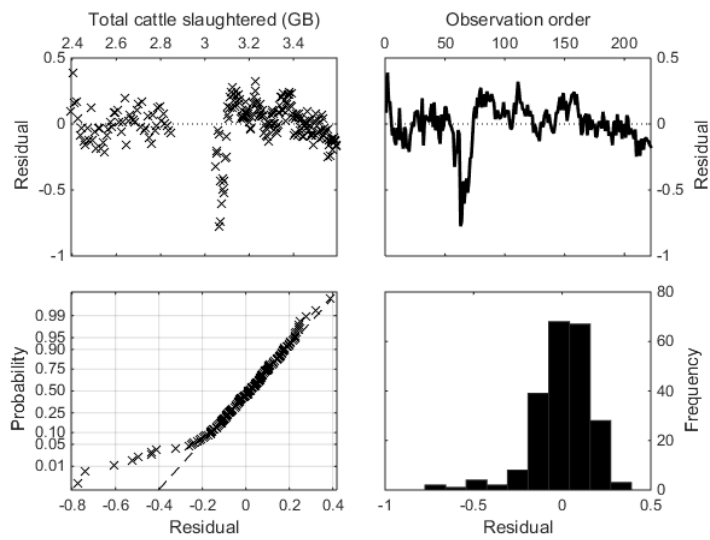

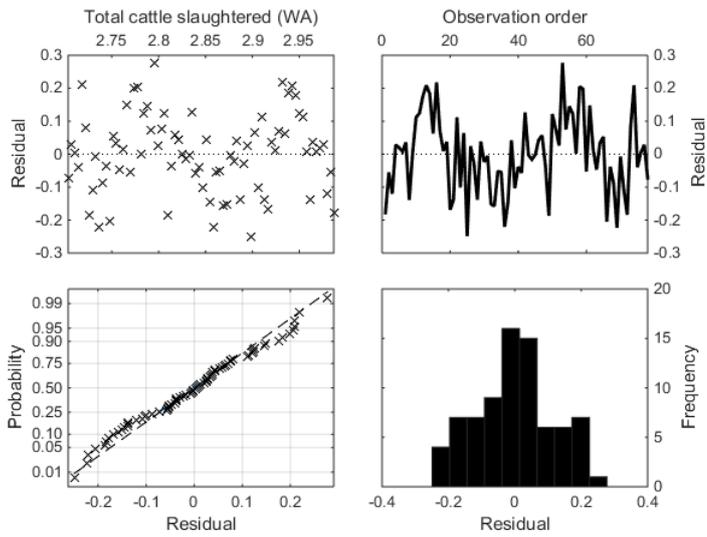

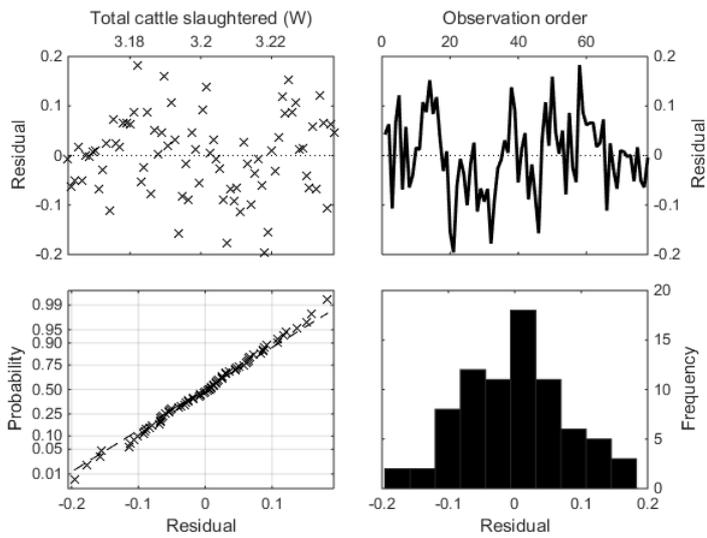

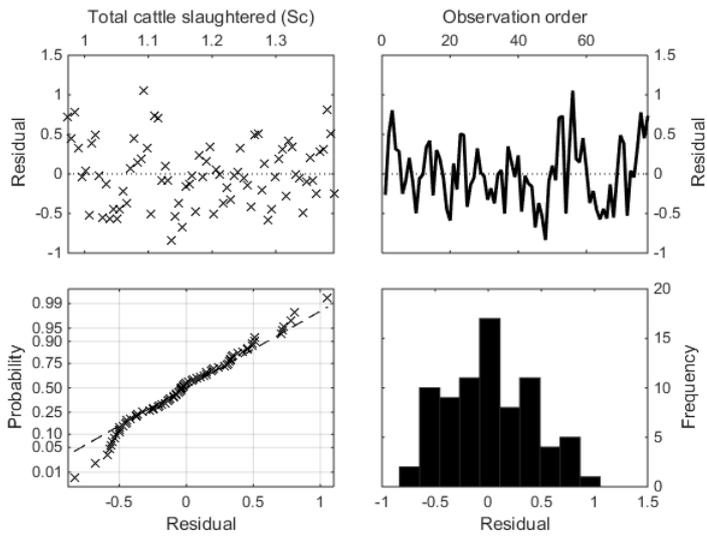

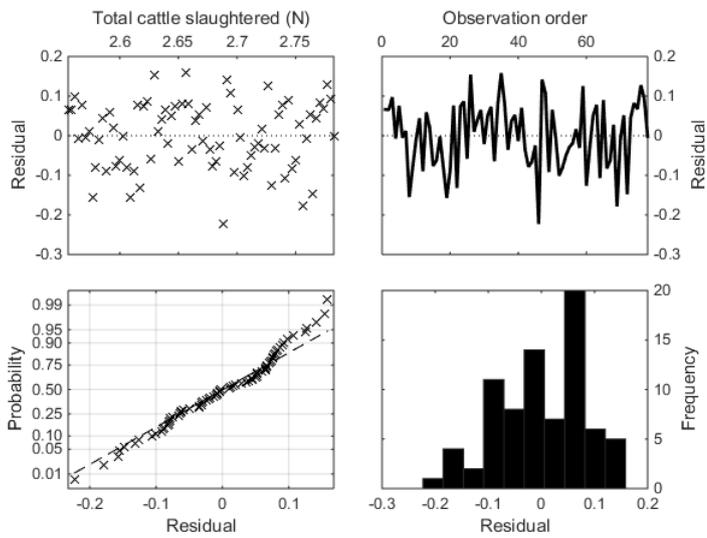

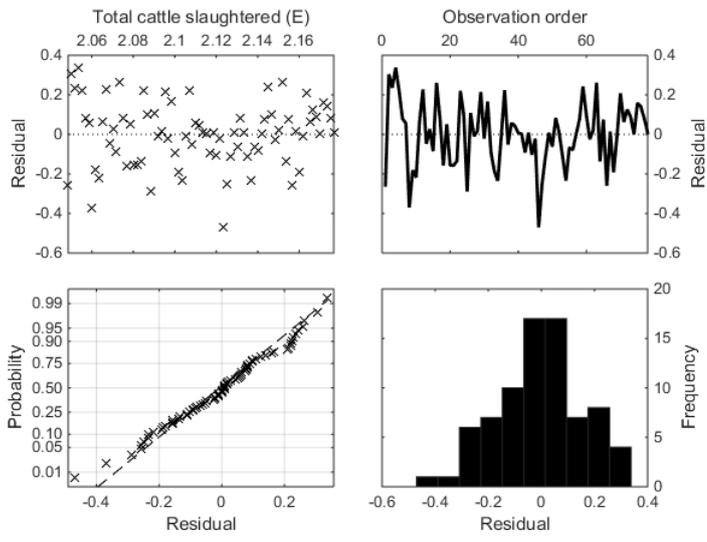

## Residuals of New herd incidents

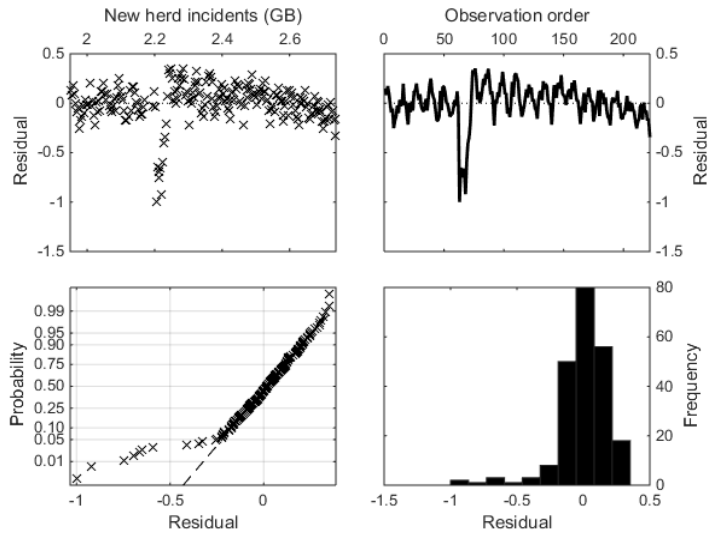

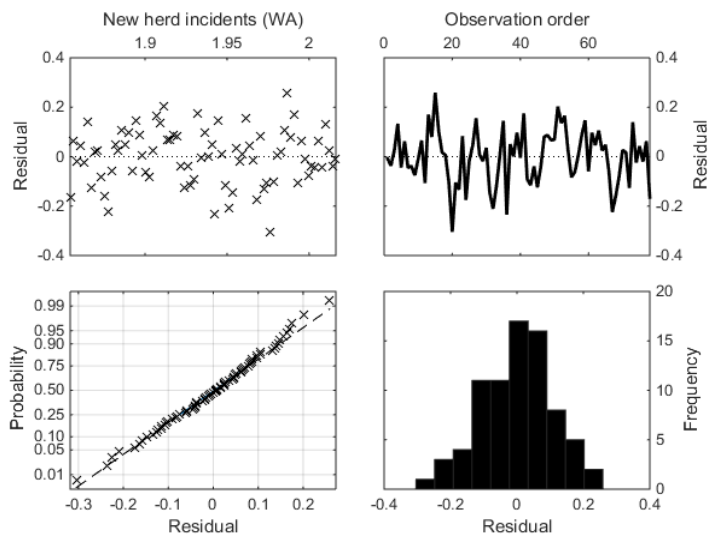

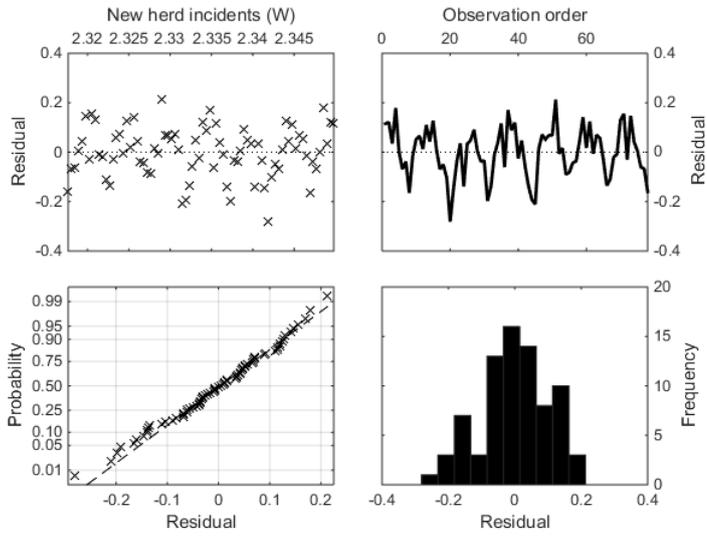

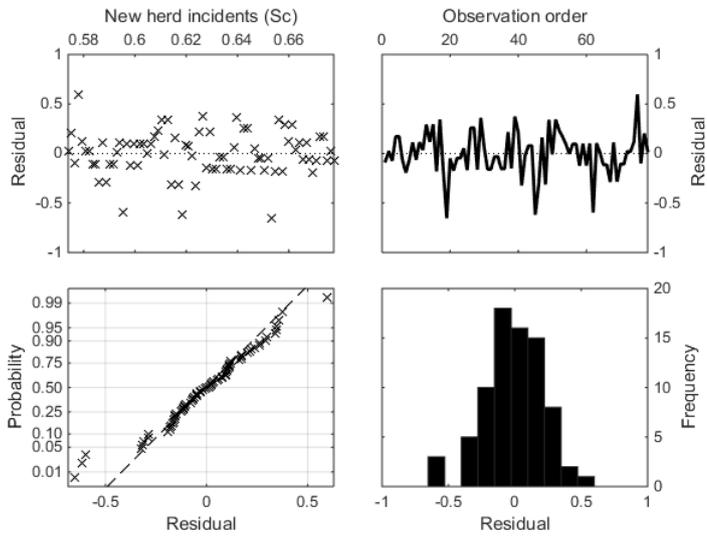

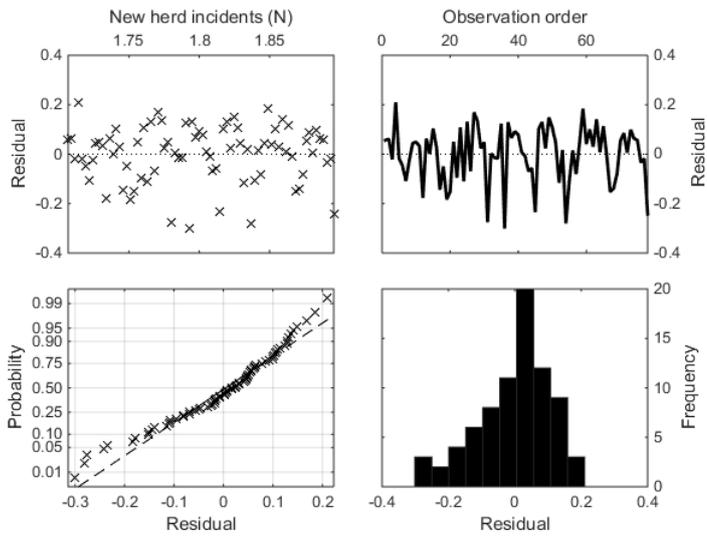

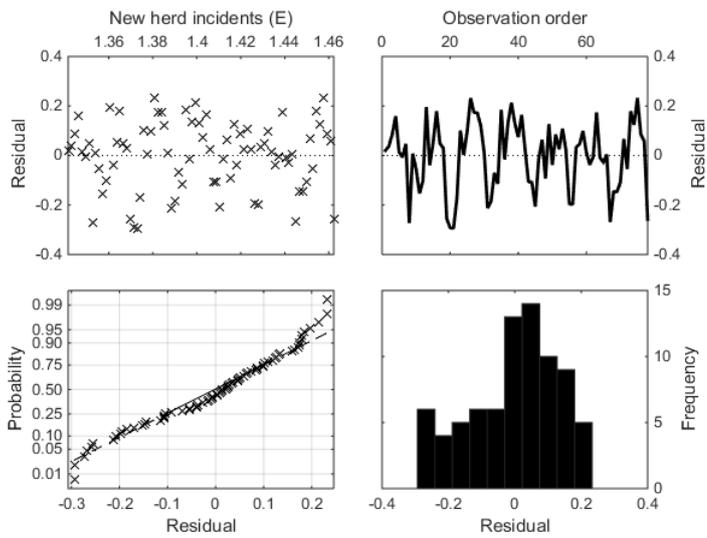

## Residuals of Herds not TB free

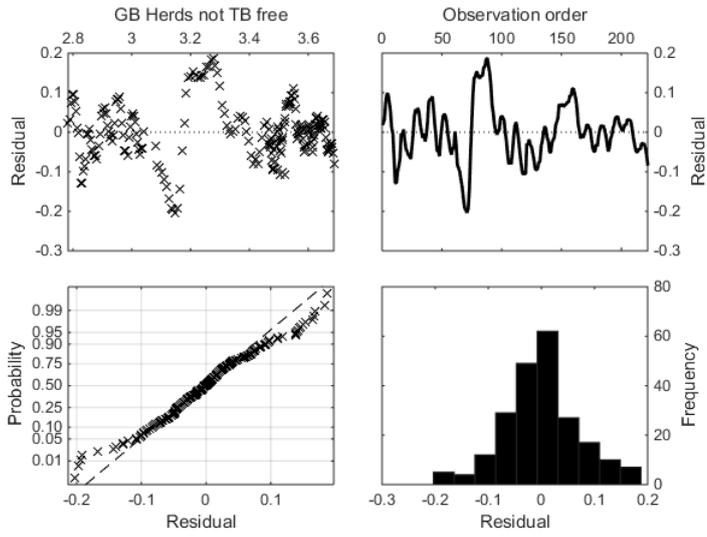

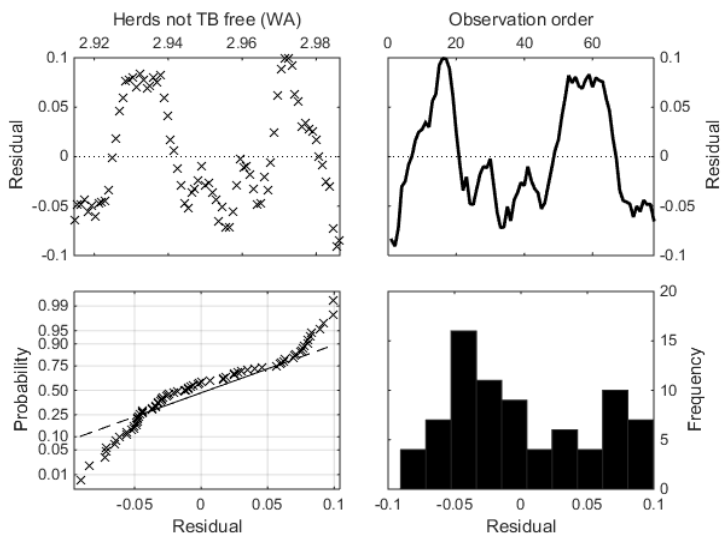

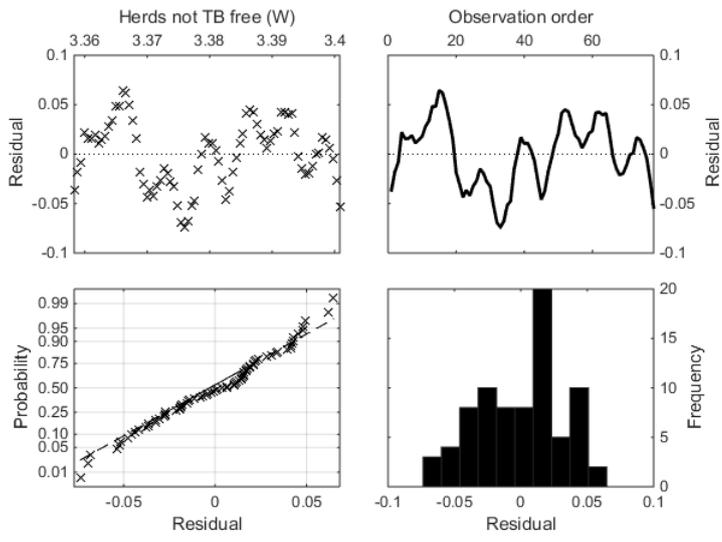

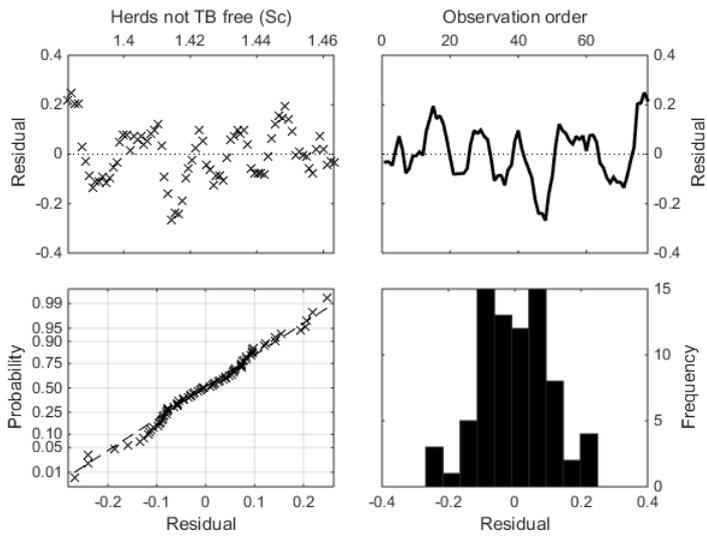

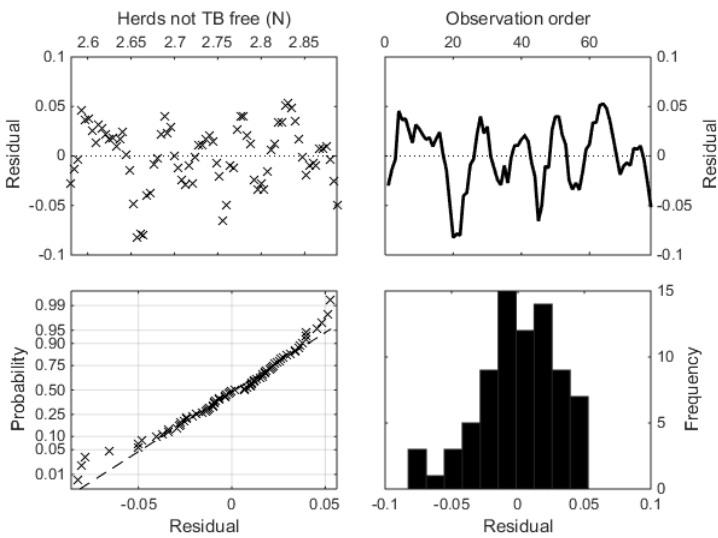

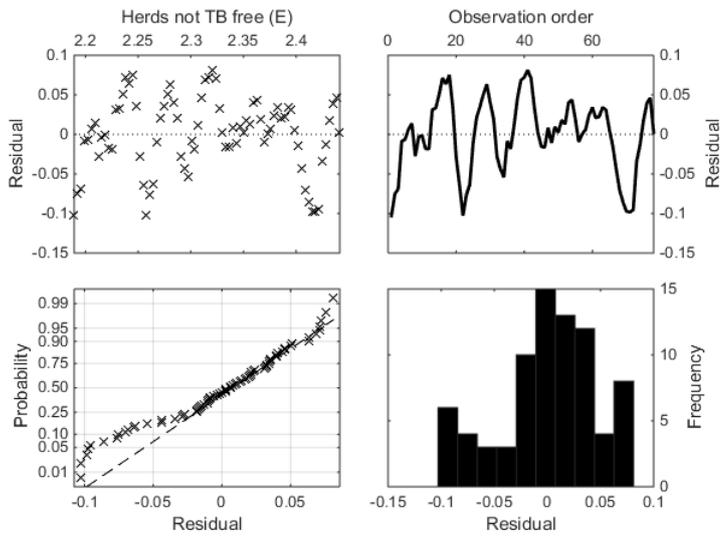

## Regression Analysis: total cattle slaughtered versus total tests herds per region

Detailed results of regression analysis of Figure 6 of the main text. In all graphs the regression is plotted with a solid red line, 95% confidence intervals of the regression with dashed green lines, and 95% prediction intervals with purple dashed lines. The original log-transformed data are plotted with blue circles. Data were $\log_{10}(x+1)$ transformed prior to the analysis. Here all regressions are plotted into the range of the data for each region for higher detail.

**Wales**

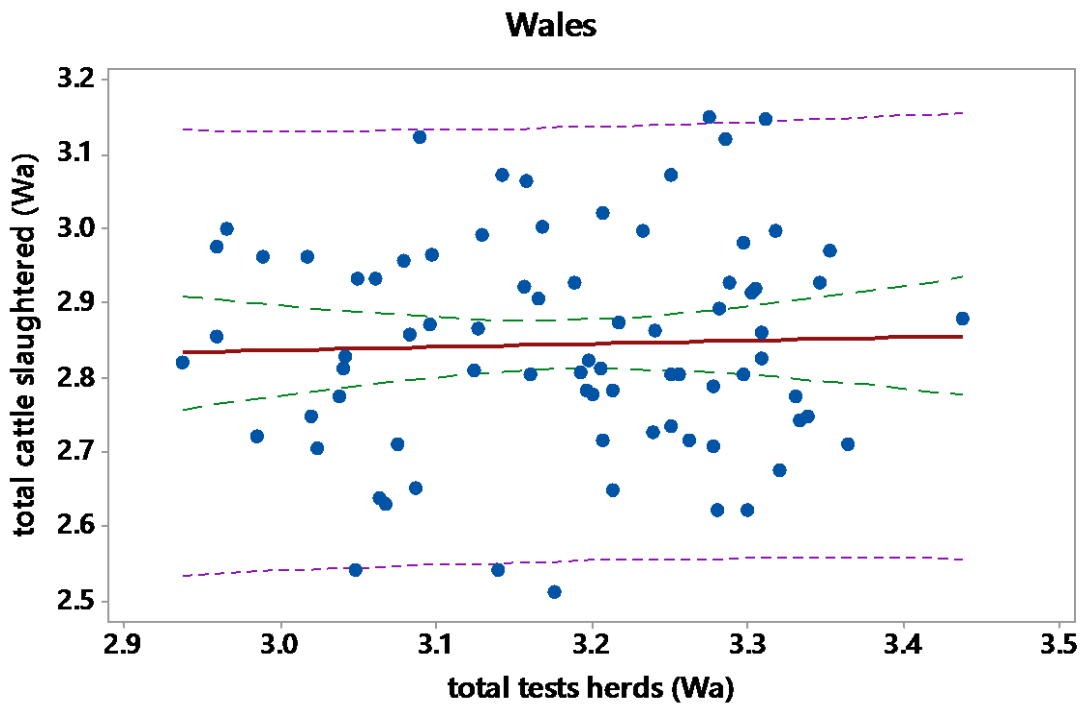

The regression equation is
total cattle slaughtered (Wa) = 2.701 + 0.0450 total tests herds (Wa)

S = 0.145855   R-Sq = 0.1%   R-Sq(adj) = 0.0%

Analysis of Variance

Source      DF   SS       MS        F     P
Regression   1  0.00216  0.0021619  0.10  0.751
Error       76  1.61679  0.0212736
Total       77  1.61895

**West England**

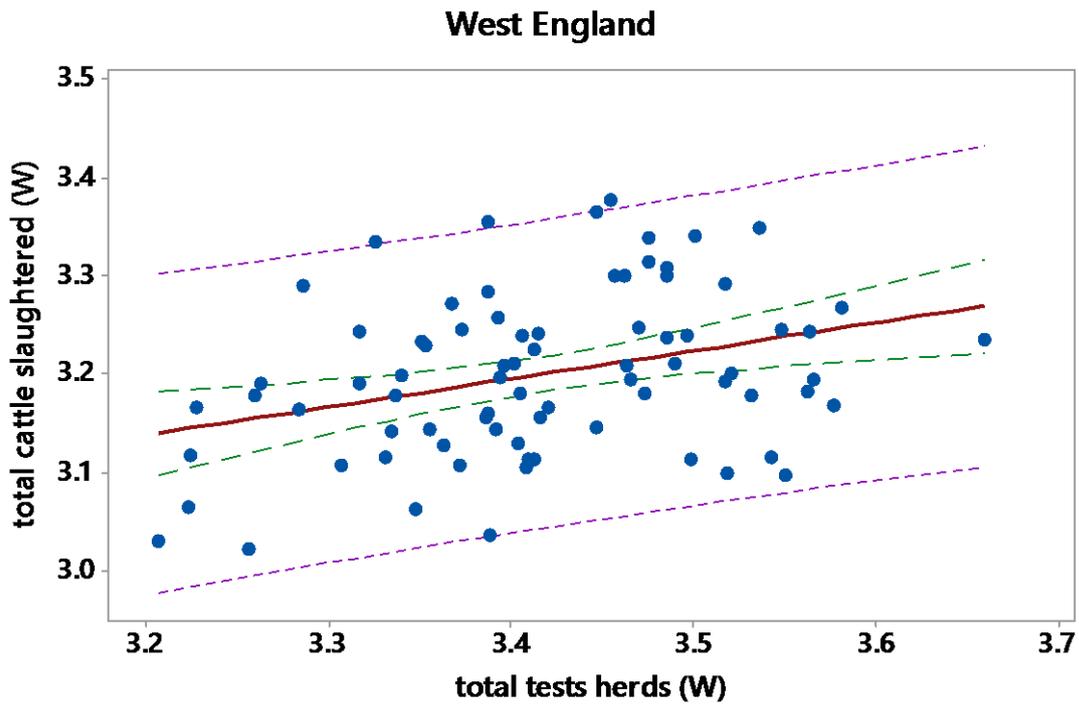

The regression equation is
total cattle slaughtered (W) = 2.222 + 0.2862 total tests herds (W)

S = 0.0783644   R-Sq = 11.3%   R-Sq(adj) = 10.1%

Analysis of Variance

| Source | DF | SS | MS | F | P |
|---|---|---|---|---|---|
| Regression | 1 | 0.059184 | 0.0591835 | 9.64 | 0.003 |
| Error | 76 | 0.466715 | 0.0061410 | | |
| Total | 77 | 0.525898 | | | |

**Scotland**

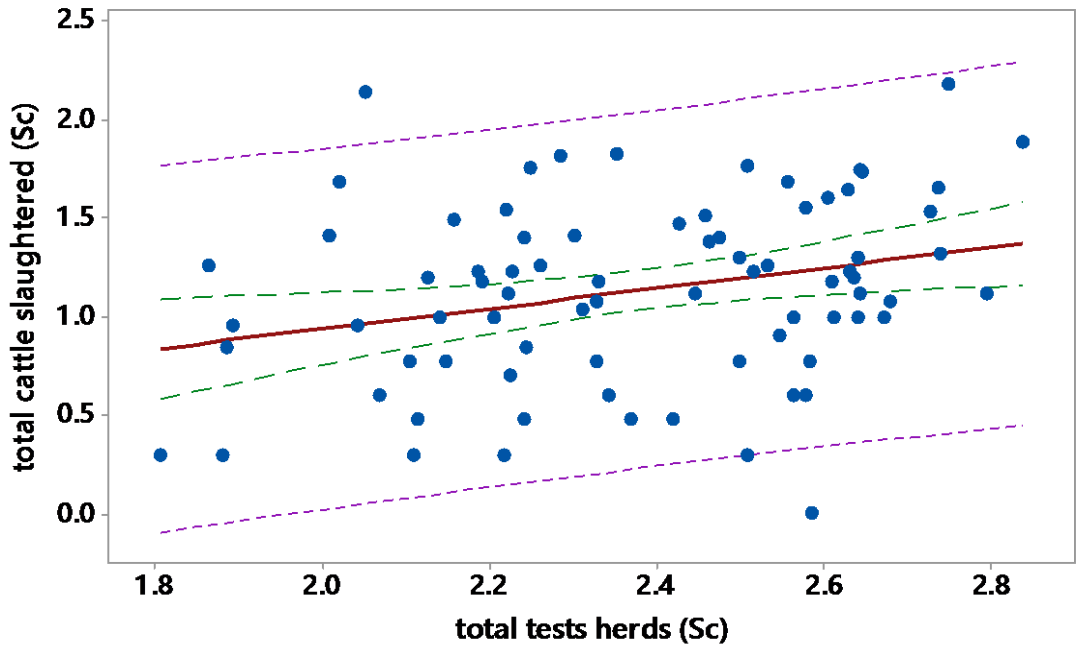

The regression equation is
total cattle slaughtered (Sc) = - 0.0968 + 0.5171 total tests herds (Sc)

S = 0.451524   R-Sq = 7.8%   R-Sq(adj) = 6.6%

Analysis of Variance

| Source | DF | SS | MS | F | P |
| --- | --- | --- | --- | --- | --- |
| Regression | 1 | 1.3079 | 1.30794 | 6.42 | 0.013 |
| Error | 76 | 15.4944 | 0.20387 | | |
| Total | 77 | 16.8024 | | | |

**North England**

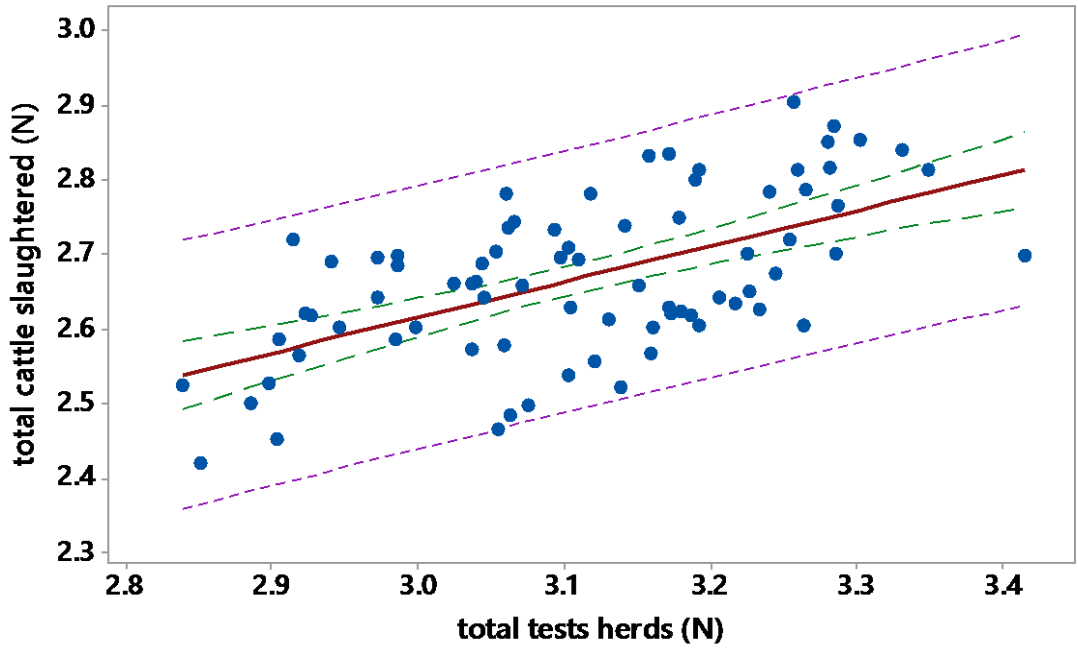

The regression equation is
total cattle slaughtered (N) = 1.178 + 0.4787 total tests herds (N)

S = 0.0877558   R-Sq = 34.1%   R-Sq(adj) = 33.3%

Analysis of Variance

| Source | DF | SS | MS | F | P |
| --- | --- | --- | --- | --- | --- |
| Regression | 1 | 0.303307 | 0.303307 | 39.39 | 0.000 |
| Error | 76 | 0.585283 | 0.007701 | | |
| Total | 77 | 0.888590 | | | |

**East England**

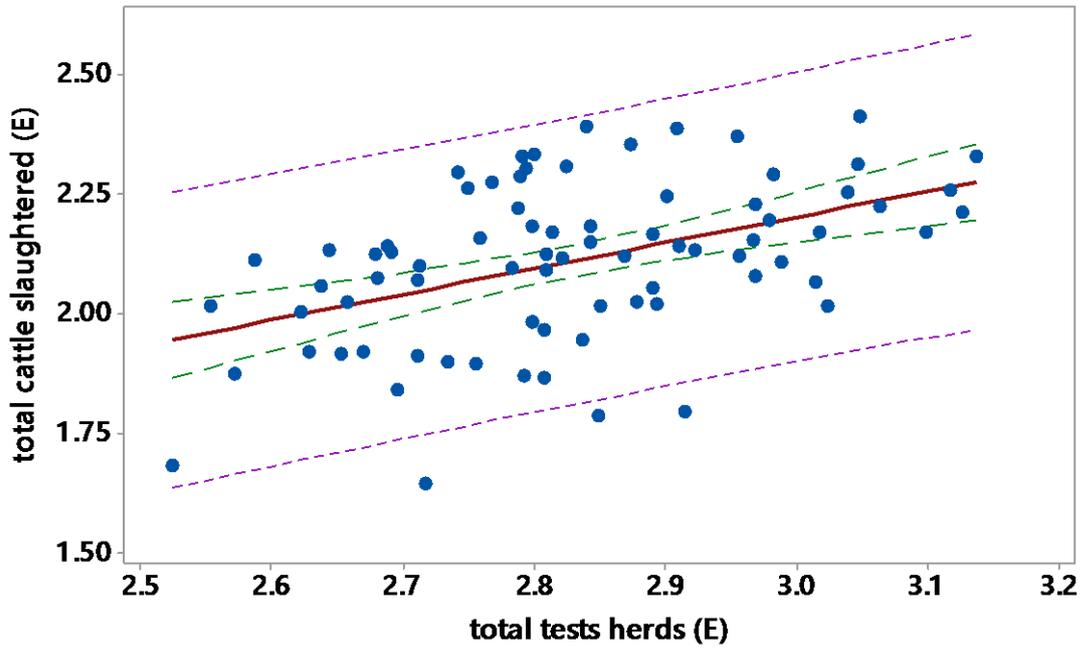

The regression equation is
total cattle slaughtered (E) = 0.5804 + 0.5400 total tests herds (E)

S = 0.149770   R-Sq = 21.4%   R-Sq(adj) = 20.4%

Analysis of Variance

| Source | DF | SS | MS | F | P |
|---|---|---|---|---|---|
| Regression | 1 | 0.46483 | 0.464834 | 20.72 | 0.000 |
| Error | 76 | 1.70476 | 0.022431 | | |
| Total | 77 | 2.16960 | | | |